\documentclass[a4paper,11pt]{article}

\pdfoutput=1
\usepackage{jcappub}

\usepackage{graphicx}
\usepackage{longtable}
\usepackage{float}
\usepackage{dcolumn}
\usepackage{bm}
\usepackage{appendix}
\usepackage{multirow}
\usepackage{color}
\usepackage[utf8]{inputenc}
\usepackage{footmisc}
\usepackage{hyperref}

\usepackage{txfonts}
\usepackage{xcolor}
\usepackage{graphicx}
\usepackage{bm}
\usepackage{ulem}
\usepackage{multirow, array, float, tabularx}
\usepackage{amsmath}
\hypersetup{
	citecolor = red,
	linkcolor = blue,
	urlcolor = magenta
}

\newcommand{\be}{\begin{equation}}
	\newcommand{\ee}{\end{equation}}
\newcommand{\bea}{\begin{eqnarray}}
	\newcommand{\eea}{\end{eqnarray}}
\newcommand{\bes}{\begin{subequations}}
	\newcommand{\ees}{\end{subequations}}

\newcommand{\bc}{\begin{center}}
	\newcommand{\ec}{\end{center}}

\usepackage{wasysym}

\begin{document}

\title{Updating constraints on phantom crossing $f(T)$ gravity}
\author[a]{F. B. M. dos Santos}\emailAdd{felipebrunomedeiros@gmail.com}

\affiliation[a]{Departamento de F\'{\i}sica,Universidade Federal do Rio Grande do Norte, Av. Sen. Salgado Filho, Natal, RN, 59078-970, Brasil}

\abstract{We establish constraints on $f(T)$ gravity by considering the possibility of a scenario that supports a phantom crossing of the equation of state parameter $\omega_{DE}$. After determining the viable parameter space of the model, while checking the impact on the background dynamics, we perform an analysis to obtain constraints on cosmological parameters and determine the viability of this scenario. To this end, we use combined data sets from cosmic chronometers (CC), baryonic acoustic oscillations (BAO), redshift space distortion (RSD) and Type Ia supernovae (SN) measurements from the latest Pantheon$+$ set, in which the impact on the absolute magnitude due to the change of the effective gravitational constant is also considered. It is found that a state where a phantom crossing of $\omega_{DE}$ happens is favored by data, and the $f(T)$ model is competitive with the $\Lambda$CDM one by statistical criteria, such as AIC and BIC. Additionally, we find evidence of the Hubble tension being alleviated within the $f(T)$ model, at the same time that it does not worsen the growth one, indicating a possibility of the present scenario as an option to address the current cosmic tensions.}

\maketitle

\section{Introduction}\label{sec1}

The success of the $\Lambda$CDM model \cite{Perlmutter:1998np, Riess:1998cb} in describing all kinds of cosmological data is constantly challenged, but the model not only is capable of being consistent with robust data sets \cite{Aghanim:2018eyx}, but it is also preferred when statistical criteria are considered. On the other hand, the standard picture presents questions that motivate the developments of a variety of extended and modified gravity scenarios \cite{Capozziello:2011et,Clifton:2011jh}. Currently, the most debated issue regarding a possible inconsistency of the $\Lambda$CDM model with cosmic history is the tension between early and late-time cosmological measurements. In particular, there is a clear discrepancy between the values of current expansion rate, given by the Hubble parameter $H_0$ measured by the early-time probes such as the one given by the Planck collaboration \cite{Aghanim:2018eyx}, of $H_0=67.4\pm0.5$ Km/s/Mpc and a variety of local determinations \cite{Riess:2019cxk,Freedman:2019jwv,Wong:2019kwg,Riess:2021jrx}. In general, a tension of more than $4\sigma$ is present, as made explicit by the most recent SH0ES measurement, of $73.3\pm1.04$ Km/s/Mpc \cite{Riess:2021jrx}. In this manner, a widely used approach to explain such discrepancy is to consider the possibility that this is the result of yet to be discovered physics, and this possibility has been extensively investigated in recent years \cite{DiValentino:2021izs}. Another issue has to do with the growth of perturbations. Measurements from Planck and cosmic shear seem to disagree, in a way that the quantity $S_8=\sigma_8\sqrt{\Omega_m/0.3}$ quantifies the growth tension, such that early universe data seem to prefer a higher rate of growth \cite{Aghanim:2018eyx,DiValentino:2020vvd}. While there is a great focus in developing theoretical models that alleviate or solve the Hubble tension, recent discussions have been considering the possibility of a given extension of the standard model as being able to alleviate both tensions at the same time \cite{Heisenberg:2022gqk,Abdalla:2022yfr}. However, it seems that if we use a parametrization of the equation of state parameter $w(z)$, such as the well-known CPL \cite{Chevallier:2000qy,Linder:2002et} model, the resolution of the Hubble tension implies a worsening of the growth one \cite{Alestas:2021xes}. On the other hand, there is an ongoing discussion on whether the $S_8$ discrepancy is severe enough for new physics to be considered \cite{Nunes:2021ipq}.

Among the many modifications of the standard model, one that is having a great deal of attention is the possibility that gravity is described by a non-zero torsion, in which the torsion scalar $T$ can be included in the field equations as a general function $f(T)$, so that we now have $f(T)$ gravity models \cite{Bengochea:2008gz,Linder:2010py,Chen:2010va,Bamba:2010wb,Nesseris:2013jea,Cai:2015emx,Krssak:2018ywd,Bahamonde:2021gfp}, developed similarly as $f(R)$ ones \cite{Sotiriou:2008rp,Nojiri:2006ri,DeFelice:2010aj}. The study of different $f(T)$ models \cite{Nunes:2016plz,Nunes:2016qyp,Xu:2018npu,Basilakos:2018arq,Anagnostopoulos:2019miu,Wang:2020zfv,Benetti:2020hxp,Briffa:2021nxg,dosSantos:2021owt} have shown that it is possible for a model to reproduce the $\Lambda$CDM one for a choice of parameters, while allowing a small deviation from the standard model, having important implications in the cosmological dynamics. The possibility of $f(T)$ gravity on alleviating cosmic tensions has also been considered in recent literature \cite{Nunes:2018xbm}. In particular, the capability of this class of models in solving both Hubble and growth tensions has been discussed in \cite{Yan:2019gbw}, in which a $f(T)$ function that would address this issue was obtained though data reconstructions. Another study focused on the use of CMB priors to investigate how well-known $f(T)$ models could address the $H_0$ tension, in which the power-law form could be a viable option by an increasing uncertainty on the deviation parameter \cite{Wang:2020zfv}. On the other hand, an exponential form was favored in this aspect when machine learning techniques are considered \cite{Aljaf:2022fbk}. In \cite{Heisenberg:2022gqk} the general requirements of a cosmological model to solve the Hubble and growth tensions are discussed, in a way that certain conditions could be established, which helps in model building. One of these conditions is the presence of a phantom crossing at some redshift close to the present time, which motivates the consideration of $f(T)$ scenarios which possess this feature, which will be done in this work. $f(T)$ models with phantom crossing were considered in earlier works \cite{Wu:2010av,Bamba:2010wb}, although some of them were shown to be in considerable tension with the standard model, according to statistical criteria \cite{Xu:2018npu}. In Ref. \cite{Awad:2017yod}, a $f(T)$ model with exponential form was introduced as an infrared (IR) correction to general relativity (GR), meaning that the effect of the parametrization will be more significant at lower redshifts, when the accelerated cosmic expansion starts. One of the main advantages of this model is that the $f(T)$ function has only one free parameter that can be fixed by the density parameters at present time, so it can be in principle compared directly to the $\Lambda$CDM model without being penalized by the information criteria, for example. A more complete analysis was done in Refs. \cite{Hashim:2020sez,Hashim:2021pkq}, at the background and linear perturbation level, where it was found that the Hubble tension can also be alleviated. Based on these considerations, in this work we consider a $f(T)$ model that can be regarded as a generalization of this scenario, where an extra free parameter is introduced, allowing the universe to undergo a phantom crossing of $\omega_{DE}$, for a given range of parameters. In this manner, after discussing the general properties of the model, we perform a numerical analysis as a way to seek for a preference of the data for a phantom crossing within the model, while also looking at the initial impact on both $H_0$ and $\sigma_8/S_8$ tensions.

This work is organized in the following manner. In Sec. \ref{sec2}, we review the $f(T)$ formalism for cosmology, while in Sec. \ref{sec3}, we present and investigate the $f(T)$ model. In Sec. \ref{sec4}, we describe the data and methodology to be used in the analysis, while in Sec. \ref{sec5}, we discuss the results. To conclude, in Sec \ref{sec6}, we present our considerations.

\section{$f(T)$ gravity and dark energy}\label{sec2}

In $f(T)$ gravity \cite{Cai:2015emx,Bahamonde:2021gfp}, the action is described by a function of the torsion scalar $T$, and reads as
\begin{equation}
	S = \frac{1}{16\pi G}\int d^4x e f(T) + S_m,
    \label{eq:2.1}	
\end{equation}
with $S_m$ describing matter fields, and $G$ is the Newtonian gravitational constant. In teleparallel gravity, the dynamics of spacetime are described by the presence of torsion, meaning that instead of the metric tensor $g_{\mu\nu}$, used for describing GR, the fundamental object is the tetrad $e_{\mu}^{A}$. This can be related with the metric tensor through  $g_{\mu\nu}=\eta_{AB}e^A_{\;\;\mu}e^B_{\;\;\nu}$, such that $e=\sqrt{-g}$, and as a consequence, the Levi-Civita connection $\bar{\Gamma}^\lambda_{\mu\nu}$ is replaced by the teleparallel one. We can get to the correspondent field equations in the following manner: Since the teleparallel connection can be related to the Riemannian one as
\begin{equation}
     {\Gamma}_{\;\;\mu\nu}^{\lambda} = \Bar{{\Gamma}}_{\;\;\mu\nu}^{\lambda} + K_{\;\;\mu\nu}^{\lambda},
    \label{eq:2.2}	    
\end{equation}
with
\begin{equation}
K^{\lambda}_{\;\;\mu\nu}\equiv \frac{1}{2}\left(T_{\mu\;\;\nu}^{\;\lambda} + T_{\nu\;\;\mu}^{\;\lambda} - T^{\lambda}_{\;\;\mu\nu}\right),
    \label{eq:2.3}	
\end{equation}
being the contortion tensor which is in turn written in terms of the torsion tensor $T_{\;\;\mu\nu}^{\lambda} = 2\Gamma^\lambda_{\;\;[\mu\nu]}$, we can contract the torsion tensor to find $R=\bar R-B+T=0$, with $T=\frac{1}{4}T^{\rho\mu\nu}T_{\rho\mu\nu} + \frac{1}{2}T^{\rho\mu\nu}T_{\nu\mu\rho} - T^{\;\;\;\;\;\rho}_{\rho\mu}T^{\nu\mu}_{\;\;\;\;\nu}$ being the torsion scalar, while $B=-\frac{2}{e}\partial_\rho\left(eT^{\mu\;\;\rho}_{\;\;\mu}\right)$ is a boundary term. In this manner, for a function $f(T)$, the field equations become \cite{Cai:2015emx,Bahamonde:2021gfp}
\begin{equation}
e^{-1}\partial_{\mu}\left(ee^{\rho}_{A}S_{\rho}^{\;\;\mu\nu}\right)f_{T} + e^{\rho}_{A}S_{\rho}^{\;\;\mu\nu}(\partial_{\mu}T)f_{TT}-f_{T}e_{A}^{\lambda}T^{\rho}_{\;\;\mu\lambda}S_{\rho}^{\;\;\nu\mu} + \frac{1}{4}e_{A}^{\nu}f(T) = 4\pi Ge^{\rho}_{A}\mathcal{T}_{\rho}^{v},
    \label{eq:2.4}	
\end{equation}
with $_T$ denoting derivatives with respect to the torsion scalar $T$, $S_\rho^{\;\;\mu\nu}\equiv\frac{1}{2}\left(K^{\mu\nu}_{\;\;\;\;\rho}+\delta^\mu _\rho T^{\alpha\nu}_{\;\;\;\;\alpha}-\delta^\nu_\rho T^{\alpha\mu}_{\;\;\;\;\alpha}\right)$ is the called super-potential, and $\mathcal{T}^\nu_\rho$ is the total energy-momentum tensor. By imposing a Friedmann-Lemaître-Robertson-Walker (FLRW) geometry as $ds^2 = dt^2 - a^2(t)\delta_{ij}dx^i dx^j$, one can derive the correspondent Friedmann equations for $f(T)$ gravity as
\begin{equation}
3H^2 = 8\pi G(\rho_{m} + \rho_{r}) - \frac{f}{2} - \frac{T}{2} + Tf_{T},
    \label{eq:2.5}	 
\end{equation}
\begin{equation}
\dot{H} = -\frac{4\pi G(\rho_{m} + P_{m} + \rho_{r} + P_{r})}{f_{T}+2Tf_{TT}},
    \label{eq:2.6}	
\end{equation}
with $H\equiv \frac{\dot{a}}{a}$ being the Hubble parameter, and $T=-6H^2$. It is clear that the choice $f(T)=T$ recovers GR, while $f(T)=T-2\Lambda$ reproduces the $\Lambda$CDM model. 

Looking at eqs. \ref{eq:2.5} and \ref{eq:2.6}, one can see that it is possible to write them in a more convenient way by defining
\begin{equation}
\rho_{DE} \equiv \frac{1}{16\pi G}\left[ 2Tf_{T}-T - f \right], \quad 
P_{DE} \equiv \frac{1}{16\pi G}\left[ \frac{f-f_{T}T + 2T^2f_{TT}}{f_{T} + 2Tf_{TT}} \right],
    \label{eq:2.7}	
\end{equation}
so that the dark energy equation of state can be written as
\begin{equation}
\omega_{de} \equiv \frac{P_{DE}}{\rho_{DE}} = -1 + \frac{(f-2Tf_{T})(f_{T}+2Tf_{TT}-1)}{(f+T-2Tf_{T})(f_{T}+2Tf_{TT})}.
    \label{eq:2.8}	
\end{equation}
We remember that the cosmological fluids considered, being non-relativistic matter, radiation and dark energy are conserved, and thus follow the usual continuity equations obtained from the conservation of the energy momentum tensor. 

It is also possible to see the effects of the model at the linear matter perturbation level. The equation for the density contrast $\delta_m\equiv\delta\rho_m/\rho_m$ for $f(T)$ gravity can be expressed as \cite{Wu:2012hs}
\begin{gather}
	\ddot{\delta}_m + 2H\dot{\delta}_m = 4\pi G_{eff}\rho_m\delta_m,
    \label{eq:2.9}	
\end{gather}
where the modification realized by the $f(T)$ function can be expressed by the presence of an effective gravitational constant $G_{eff}=G/f_T$, such that a reasonable deviation from the Newtonian one can be achieved, depending on the model considered. To investigate the impact of the model in the growth of matter perturbations by comparison with data, we can use the measurable quantity $f\sigma_8$
\begin{gather}
	f\sigma_8\equiv \sigma_8\frac{\delta_m'}{\delta_{m,0}},
    \label{eq:2.10}	
\end{gather}
with $\sigma_8$ being the deviation from $\delta_m$ in spheres of radius $8h^{-1}$Mpc, and the prime denotes a derivative with respect to $N=\log a$. We can solve eq. \ref{eq:2.9} by assuming initial conditions at sub-horizon scales in the matter-dominated era, such that we can obtain restrictions on the $\sigma_8$ parameter from observations. 

\section{Phantom crossing exponential $f(T)$ model}\label{sec3}

In this work, we will consider the following form of $f(T)$ function
\begin{equation}
f(T) = Te^{\alpha\left(\frac{T_0}{T}\right)^b}.
    \label{eq:3.1}	
\end{equation}
This function is dependent on the parameters $\alpha$ and $b$, where the latter generalizes the exponential function $e^{\alpha T_0/T}$; the model $f(T)=Te^{\alpha T_0/T}$ was discussed in \cite{Awad:2017yod} and further investigated in \cite{Hashim:2020sez,Hashim:2021pkq} for the background plus linear perturbation level. This particular model, characterized by the $b=1$ case in eq. \ref{eq:3.1}, has the advantage of not adding new extra degrees of freedom to the usual cosmological scenario, in this way, one is able to compare this model and the $\Lambda$CDM in an equal manner. Another characteristic is that this form does not reduce to the standard model for a choice of parameters, although a close to $\Lambda$CDM evolution is achieved at higher redshifts, meaning that the change in dynamics imposed by the $f(T)$ function is more relevant at late times. 

Similarly, the function in eq. \ref{eq:3.1} is not capable of reproducing the $\Lambda$CDM model for a choice of $b$, which introduces a deformation in the argument of the exponential term; thus, we are able to control how this term evolves as the redshift increases. We will now see the consequence of a changing $b$. The energy density and pressure are given by
\begin{gather}
	\rho_{DE} = \frac{3H_0^2E^2}{8\pi G} \left\{1 - \left[ 1 - \frac{2\alpha b}{E^{2b}}\right]e^{\alpha/E^{2b}}\right\}, \\\nonumber
	P_{DE} = -\frac{3H_0^2E^2}{8\pi G} \frac{\alpha b(2\alpha b + (2b - 1)E^{2b})}{(E^{4b}+\alpha b (2\alpha b + (2b -3)E^{2b}))}
    \label{eq:3.2}	
\end{gather}
with $E(z)\equiv H(z)/H_0$, being the normalized Hubble parameter. Using eq. \ref{eq:2.5}, we find that the Friedmann equation in terms of redshift $z$ for this model is
\begin{equation}
E^2(z)e^{\alpha/E^{2b}(z)}\left(1-\frac{2b\alpha}{E^{2b}(z)}\right)=\Omega_{m0}(1+z)^3 + \Omega_{r0}(1+z)^4,
\label{eq:3.3}
\end{equation}
with $\Omega_{i,0}\equiv\frac{8\pi G\rho_{i,0}}{3H_0^2}$. The relation between $b$ and $\alpha$ is found by setting $z=0$ ($E=1$), so that we find
\begin{equation}
e^{\alpha}\left(1-2b\alpha\right)=\Omega_{m0} + \Omega_{r0},
\label{eq:3.4}
\end{equation}
from which an exact expression can be found as
\begin{equation}
	\alpha = \mathcal{W}\left(-\frac{(\Omega_{m0}+\Omega_{r0})}{2be^{\frac{1}{2b}}}\right) + \frac{1}{2b},
	\label{3.5}
\end{equation}
with $\mathcal{W}$ being the Lambert function. Once more, as the choice $b=1$ reproduces the IR model discussed in \cite{Awad:2017yod,Hashim:2020sez,Hashim:2021pkq}, we can immediately investigate the behavior of the model at background level, as the parameter $b$ is changed.

\begin{figure*}
	\centering
	\includegraphics[width=0.49\columnwidth]{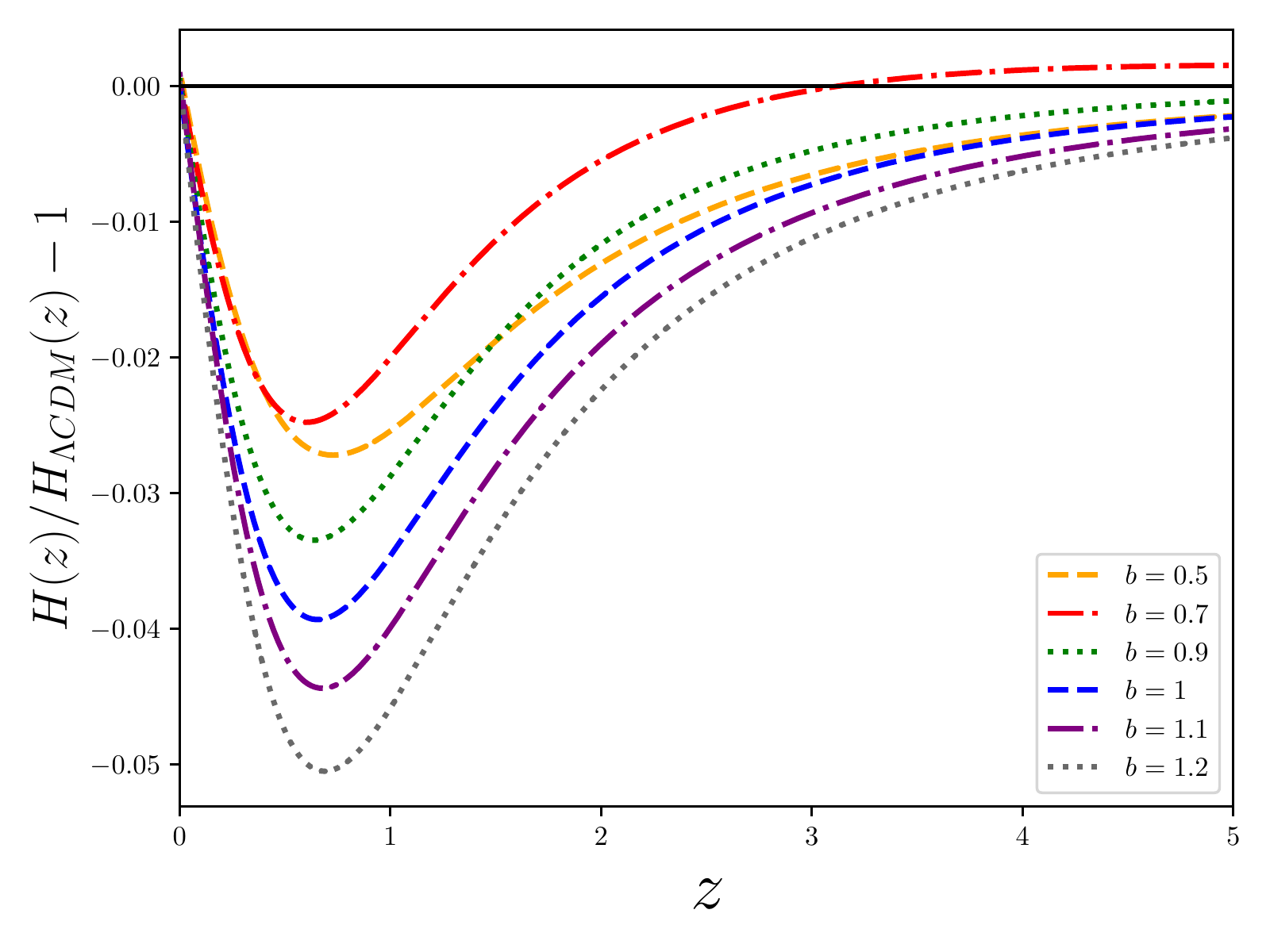}
	\includegraphics[width=0.49\columnwidth]{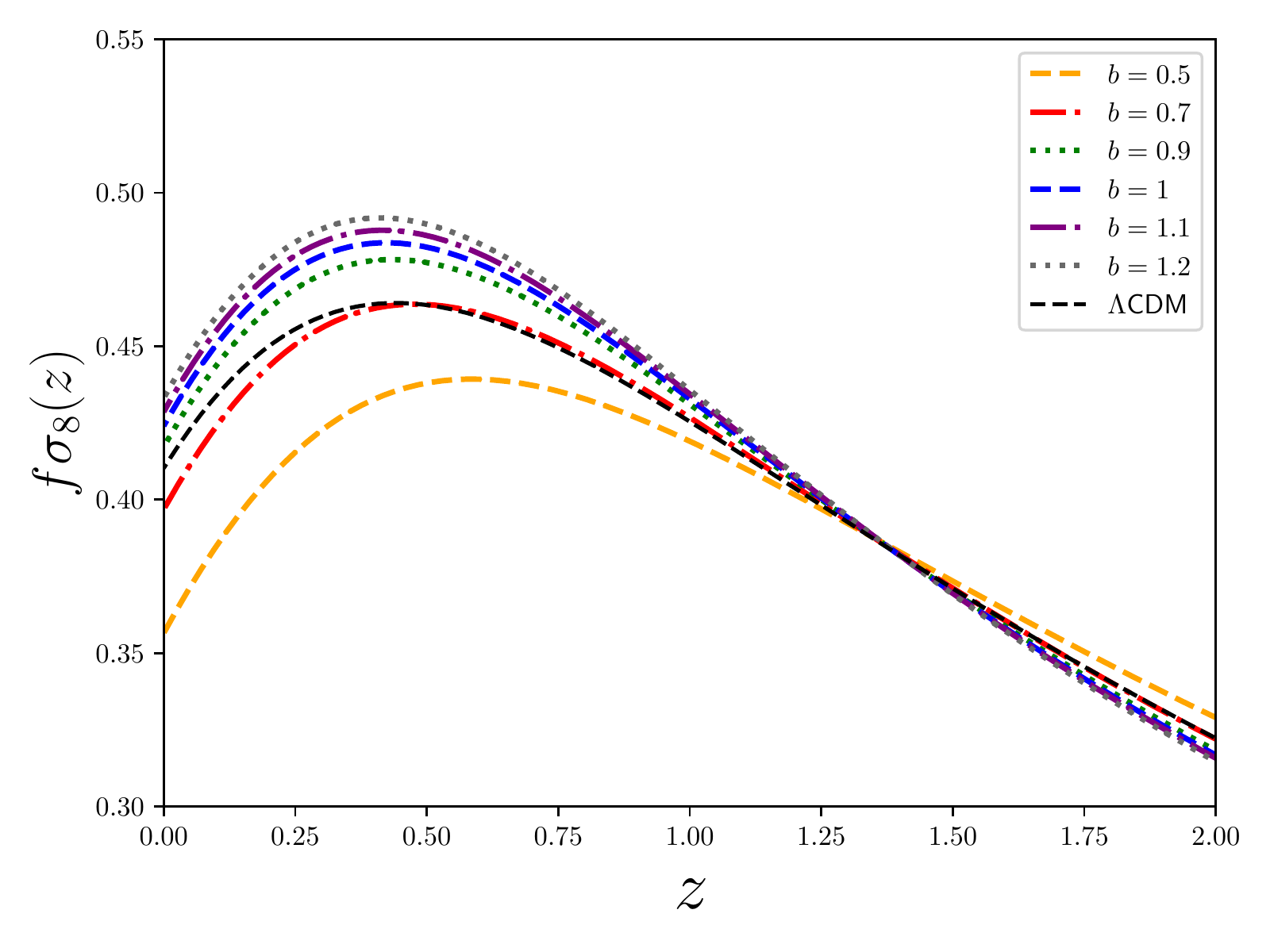}
	\includegraphics[width=0.49\columnwidth]{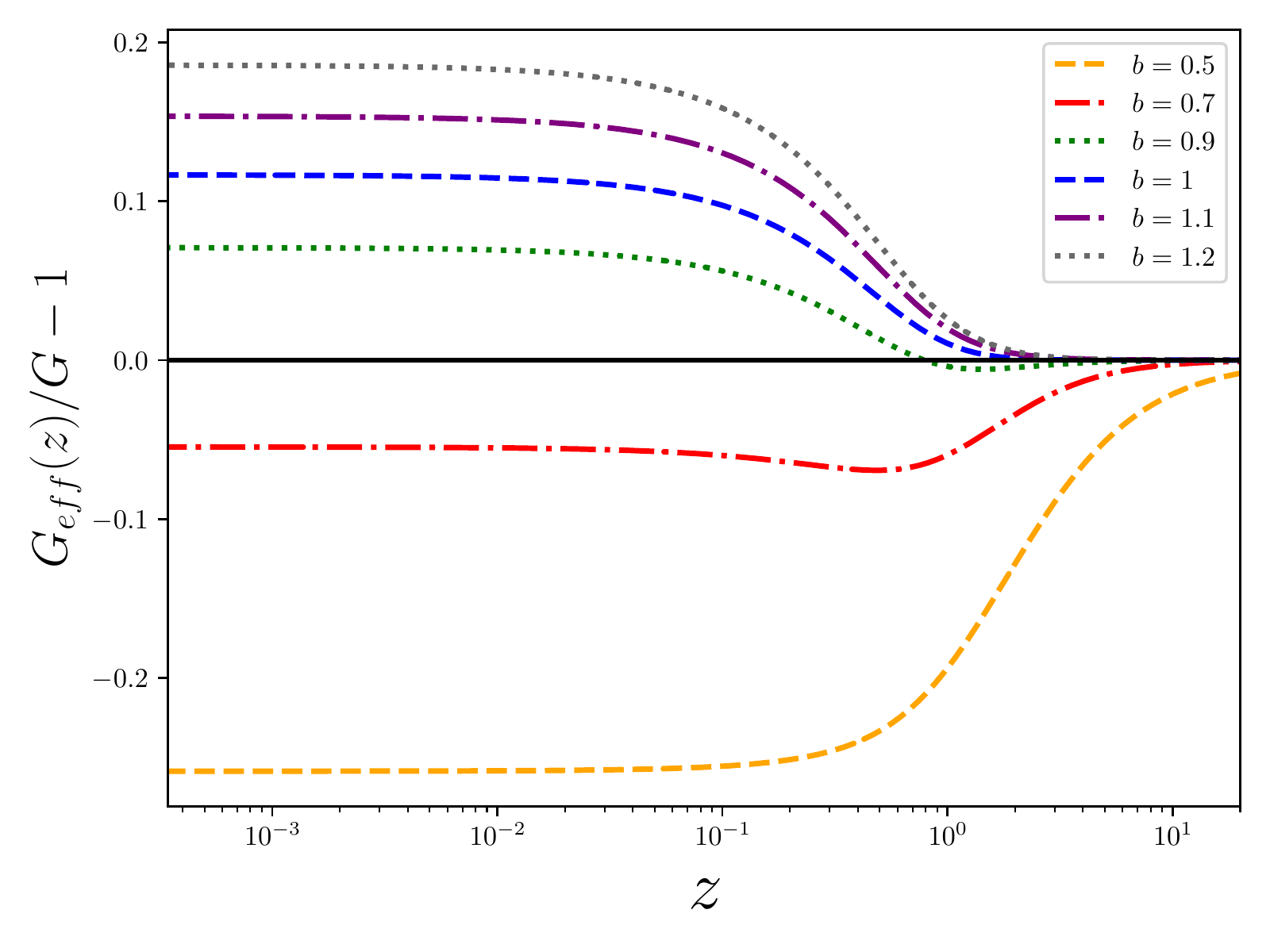}				
	\caption{The change in the Hubble parameter $H(z)/H_{\Lambda CDM}-1$(upper left panel), and the $f\sigma_8$ quantity (eq. \ref{eq:2.10}) (upper right panel) both as a function of redshift for different $b$, where we have fixed $\sigma_{8,0}=0.8, \Omega_{m0}=0.3, H_0=70$ Km/s/Mpc. The lower panel shows the effective Newtonian gravitational constant as $G_{eff}/G-1$, also as a function of $z$. The $\Lambda$CDM model is expressed by the black line in all plots.  }
	\label{fig:1}
\end{figure*}

The impact of the parameter $b$ in cosmological dynamics can be seen in fig. \ref{fig:1}, where we plot the evolution of the Hubble parameter for the generalized exponential model (upper left panel), as the fractional change with respect to the $\Lambda$CDM one. We consider a significant range of $b$ around $b=1$, and the effect of changing $b$ is the following: For $b>1$ the curves are pushed toward lower $H(z)$, while as $b$ decreases, the predictions seem to lean towards the $\Lambda$CDM curve, as it is noticeable by the red curve ($b=0.7$), which tends to approach the $\Lambda$CDM solution faster as $z$ increases. However, a closer inspection indicates that as $b$ decreases further, the values for $H(z)$ start going to lower values again, confirming the impossibility of the model to reproduce the $\Lambda$CDM one for a choice of $b$.

We can also see, from eq. \ref{eq:3.3}, that the Friedman equation can be rewritten as 
\begin{equation}
	E^2(z)=\Omega_{m0}(1+z)^3 + \Omega_{r0}(1+z)^4 + \Omega_{DE,0}y(z,b,\alpha),
	\label{eq:3.6}
\end{equation}
with $\Omega_{DE,0}=1-\Omega_{m0}-\Omega_{r0}$, and
\begin{equation}
	y(z,b,\alpha)\equiv \frac{E^2(z)}{\Omega_{DE,0}}\left[1-e^{\alpha/E^{2b}(z)}\left(1-\frac{2b\alpha}{E^{2b}(z)}\right)\right]
	\label{eq:3.7}
\end{equation}
being the distortion function already introduced in previous works in $f(T)$ gravity. At this moment, it is important to discuss some limits on the $b$ parameter. By inspecting the function given by eq. \ref{eq:3.7}, we notice that there might be a value of parameter for which $y$ is negative, implying a negative $\Omega_{DE}$. We want to avoid this behavior by checking the conditions for which this might happen. In particular, a negative $y$ is possible if $1-e^{\alpha/E^{2b}(z)}\left(1-\frac{2b\alpha}{E^{2b}(z)}\right)<0$, so we can determine, for a given $b$, what is the value of $E$ that satisfies the said condition. We find that $y$ is zero when
\begin{equation}
	E = \left[\frac{2b\alpha}{2b\left[\mathcal{W}\left(-\frac{1}{b e^{\frac{1}{2b}}}\right) + 1\right]}\right]^{1/b},
	\label{eq:3.8}
\end{equation}
for which a real solution only exists when $b<1/2$. This means that in this limit, at some point in time, $\Omega_{DE}$ will be negative, and that is why we set $b=1/2$ as our lower limit for our plots, but allow such values in the subsequent statistical analysis to be discussed in sections \ref{sec4} and \ref{sec5}, in order to see what behavior is indicated by cosmological data. As for other limits of this model, we note that first, the GR limit of this scenario is still achieved for $\alpha=0$, regardless of $b$. Second, in the distant future (i.e. $z\rightarrow-1$), we find from eq. \ref{eq:3.3} that $E_{dS}=\left(2\alpha b\right)^{\frac{1}{2b}}$, representing a de Sitter solution, again reproducing the result from the pure exponential model for $b=1$.

The behavior of $f\sigma_8$ is also shown in fig. \ref{fig:1} (upper right panel). It is possible to see that as $b$ grows, a larger growth of matter fluctuations is predicted, in comparison with the $\Lambda$CDM model (black line). Lower $b$, as represented by the orange line ($b=0.5$) seem to be disfavored, as we note a significant decrease of $f\sigma_8$ at lower redshifts. We also see that the case $b=0.7$ (red line) provides a similar fit as the $\Lambda$CDM model. Also in fig. \ref{fig:1}, on the lower panel, we show the ratio $G_{eff}/G-1=1/f_T-1$. One can note that is possible to achieve $G_{eff}=G$ for all values of $b$ considered at high $z$; however, for increasing $b$, a sudden increase around $z\sim 1$ is seen, in a way that we would have a larger effective gravitational constant today, in contrast with the lower $b$ case, well represented by the orange curve ($b=0.5$).

We then check the main feature of the model by looking at the equation of state parameter $\omega_{DE}$, given by
\begin{equation}
	\omega_{DE}(z) = \frac{\alpha b\left[2\alpha b + (2b - 1)E^{2b}(z)\right]}{\left[\left( 1 - \frac{2\alpha b}{E^{2b}(z)}\right)e^{\alpha/E^{2b}(z)} - 1\right]\left[E^{4b}(z)+\alpha b (2\alpha b + (2b -3)E^{2b}(z))\right]}, 	
	\label{eq:3.9}
\end{equation}

\begin{figure*}
	\centering
	\includegraphics[width=0.49\columnwidth]{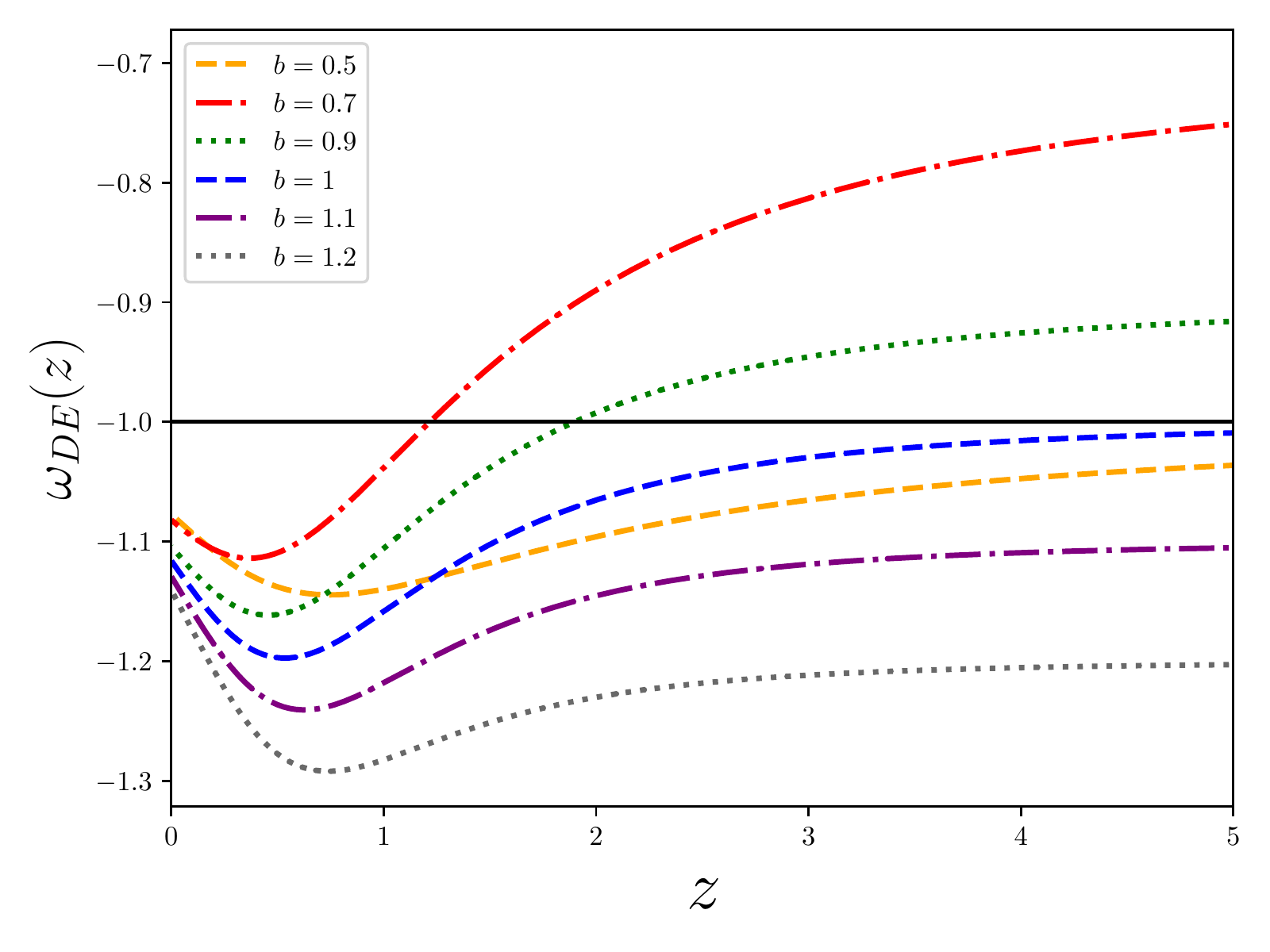}
	\includegraphics[width=0.49\columnwidth]{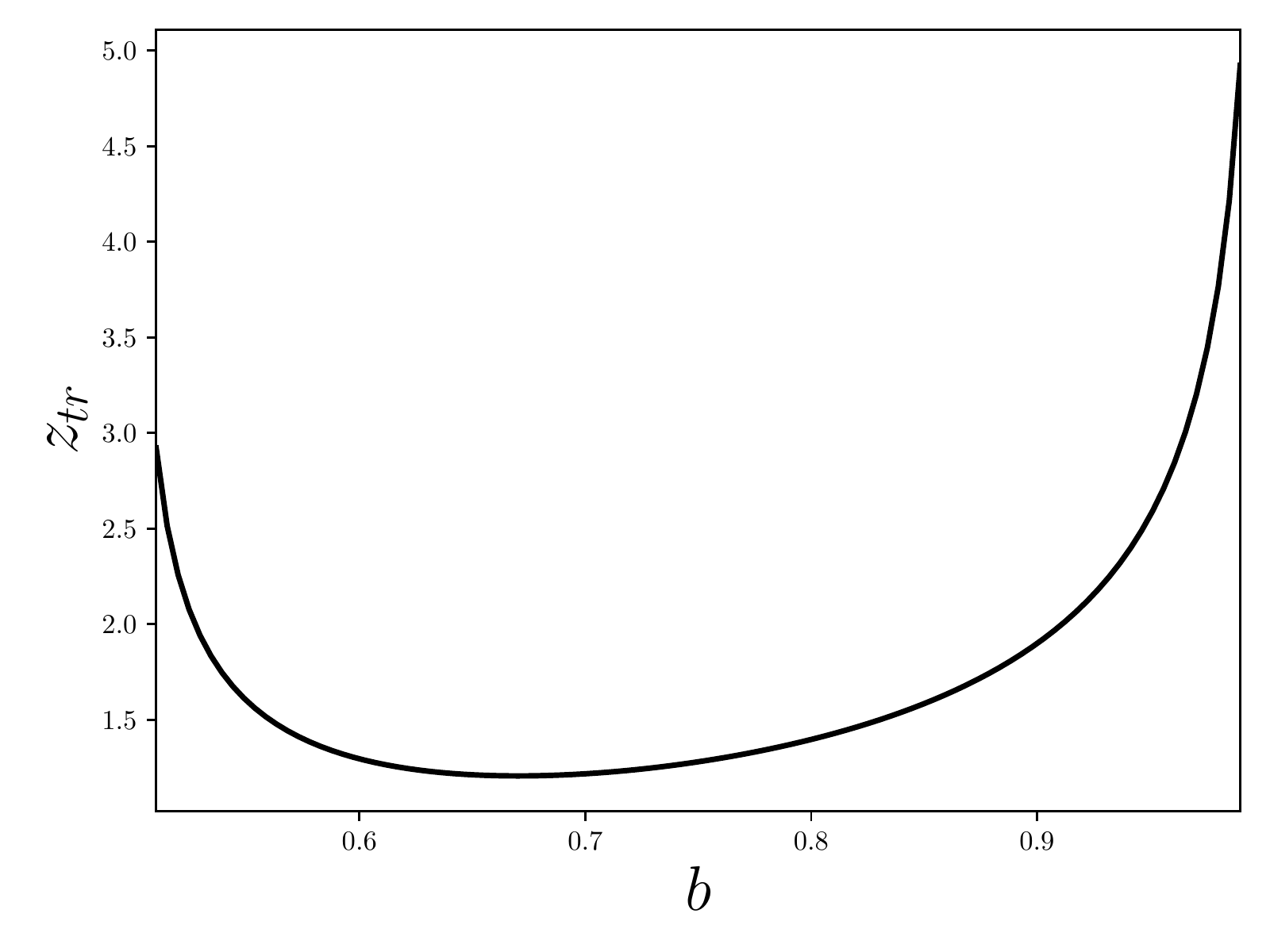}	
	\caption{The equation of state parameter $\omega_{DE}$ for the $f(T)$ model (left panel), and the redshift $z_{tr}$ for which the phantom crossing happens (right panel). We have fixed $\Omega_{m0}=0.3,H_0=70$ Km/s/Mpc.}
	\label{fig:2}
\end{figure*}

In fig. \ref{fig:2} (left panel), we show $\omega_{DE}$, as a function of redshift. The extra parameter $b$ controls the value of $\omega_{DE}$ at present times, and also provides the possibility of a phantom crossing, for some selected values. The range of $b$ for which this happens and the correspondent redshift can be computed by setting $\omega_{DE}=-1$ in eq. (\ref{eq:3.9}) for which there is always a solution corresponding to the distant future, being $E_{dS}=\left(2\alpha b\right)^{\frac{1}{2b}}$, but also another one corresponding to the phantom crossing. Restricting our search from $E=1$, we can find numerically the value of $E$ for which the transition to a phantom state happens, and then substitute the result into eq. (\ref{eq:3.3}) to estimate the transition redshift $z_{tr}$. We find that the transition only happens in the limit $1/2<z_{tr}<1$, as also seen in fig. \ref{fig:2} (right panel), where $z_{tr}$ is shown as a function of $b$. We clearly see the increase in the curve when $b$ reaches one of the limits, as for both $b=1/2$ and $b=1$, $\omega_{DE}$ goes towards $\omega_{DE}=-1$ at high $z$. For values of $b$ higher than one, however, we see that the Universe is always in a state characterized by a phantom equation of state parameter.

Another parallel between the model in eq. (\ref{eq:3.1}) and a previous work in the literature can be made. In \cite{El-Zant:2018bsc} a model with a form $f(T)=T + \alpha\frac{T_0^{1+n}}{T^n}$ was studied. The term $\alpha\frac{T_0^{1+n}}{T^n}$ introduces a correction that becomes most relevant at small $T$, at the same time that it reproduces a cosmological constant for $n=0$. By noting that at high $T$, the model in eq. (\ref{eq:3.1}) can be approximated as $f(T)=T+\alpha\frac{T_0^b}{T^b}$, we expect that the generalized exponential model would reproduce the one in \cite{El-Zant:2018bsc} for high-$z$, by making the correspondence $b=1+n$. This indeed happens, as $\omega_{DE}$ tends to a constant depending on $n$ at higher redshifts which is exactly the behavior described in \cite{El-Zant:2018bsc}. However, the main difference here is the low-$z$ behavior. Only when the $f(T)$ function is written in the form given by eq. (\ref{eq:3.1}), the phantom crossing behavior can be present, so in the lower redshift regime, this correspondence between models breaks down, and they become two different classes of models.

\section{Observational data and methods}\label{sec4}

To check the viability of these models, we will perform a statistical analysis using the Monte Carlo Markov Chain (MCMC) method, where we compare the predictions of both $\Lambda$CDM and $f(T)$ models with different data sets of the cosmological observables.

\begin{itemize}

\item Hubble parameter (CC)

We include measurements of the Hubble parameter $H(z)$ obtained from cosmic chronometers \cite{Jimenez:2001gg}. The data set is composed by 31 points, in the redshift range of $0.07\leq z\leq$ 1.965, from which 16 uncorrelated measurements are taken from \cite{Zhang:2012mp,Simon:2004tf,Ratsimbazafy:2017vga,Stern:2009ep}, together with 15 correlated ones, as discussed in \cite{Moresco:2012jh,Moresco:2015cya,Moresco:2016mzx,Moresco:2018xdr,Moresco:2020fbm}, such that we include the contribution of the respective covariance matrix. \footnote{Available at \url{https://gitlab.com/mmoresco/CCcovariance/}}

\item Baryonic acoustic oscillations (BAO)

Baryonic Acoustic Oscillations data is also considered; we use measurements from 6dFGS \cite{Beutler:2011hx}, with $r_s(z_d)/D_V(z_{eff}=0.106) = 0.336 \pm 0.015$ and SDSS DR7 \cite{Ross:2014qpa}, where $D_V(z_{eff} = 0.15) = (664 \pm 25)(r_d/r_{d,fid})$, and fiducial value $r_{d,fid}=148.69$ Mpc. $r_d$ is the value of the sound horizon at the drag epoch $z_d$, where
\begin{gather}
	r_d = \int_{z_d}^{\infty}\frac{c_s(z')}{H(z')}dz', \quad c_s(z)=\frac{c}{\sqrt{3\left(1+\frac{3\Omega_{b,0}}{4\Omega_{\gamma,0}(1+z)}\right)}},
	\label{eq:4.1}
\end{gather}
while $D_V(z)=\left(\frac{cz}{H(z)}\frac{D_L^2(z)}{(1+z)^2}\right)^{1/3}$ is the volume-averaged distance, and $D_L=c(1+z)\int_{0}^{z}\frac{dz'}{H(z')}$ is the luminosity distance. We also take $\Omega_{\gamma,0}=2.469\times 10^{-5}/h^2$. Furthermore, we also consider data from BOSS-DR12 \cite{BOSS:2016wmc} and WiggleZ \cite{Kazin:2014qga}, in which the following quantities are measured
\begin{gather}
	D_M(z)\frac{r_{s,fid}}{r_s(z_d)}, \quad H(z)\frac{r_s(z_d)}{r_{s,fid}}, \quad \text{\cite{BOSS:2016wmc}}
	\label{eq:4.2}
\end{gather}
and
\begin{gather}
	D_V(z)\frac{r_{s,fid}}{r_s(z_d)}, \quad \text{\cite{Kazin:2014qga}}
	\label{eq:4.3}
\end{gather}
with $D_M(z)=D_L/(1+z)$. This makes a total of 11 measurements, where we also note that the data points from BOSS-DR12 and WiggleZ are correlated, so we need to take this into account.

\item Type Ia Supernovae (SN)

To help constraining parameters of the model, we add data from the Type Ia supernovae from the Pantheon$+$ compilation \cite{Scolnic:2021amr,Brout:2022vxf}\footnote{Available at \url{https://pantheonplussh0es.github.io/}}, an updated sample from the previous Pantheon one \cite{Scolnic:2017caz}. It is composed of measurements from 1701 supernovae light curves in the redshift range $0.001<z<2.26$. As 77 of these curves are derived from galaxies that host Cepheids, from the SH0ES collaboration, one needs to deal with them properly. We choose to work only with non-Cepheid sources, so the data set is left with 1624 points. As the observable is the distance modulus, defined as $\mu=m_B-\mathcal{M}_B$, with $m_B$ being the observed apparent magnitude, while $\mathcal{M}_B$ is the absolute magnitude which is treated as a nuisance parameter in the statistical analysis, we can then determine the difference vector as $\mu_i-\mu_{th}(z)$, where
  \begin{gather}
  	\mu_{th}(z) = 5\log_{10}D_L(z) + 25.
  	\label{eq:4.4}
  \end{gather}
While the parameter $\mathcal{M}_B$ is taken as a constant, it is argued that a change in the value of Newton's constant $G$ could affect $\mathcal{M}_B$ due to the dependence of the absolute luminosity on the Chandrasekhar mass, which introduces a scaling as $L\sim G^{-3/2}$ \cite{Amendola:1999vu,Garcia-Berro:1999cwy,Gaztanaga:2001fh,Nesseris:2006jc}. In this manner, if a modified gravity model introduces a change in the gravitational constant, it is possible to estimate the correction to the distance modulus $\mu$. In the context of $f(T)$ gravity this has been recently done in \cite{Kumar:2022nvf}, so we also take this argument in consideration for the model studied in this work by following this approach, so the distance modulus is now expressed as

\begin{gather}
	\mu_{th}(z) = 5\log_{10}D_L(z) + 25 + \frac{15}{4}\log_{10}\frac{G_{eff}(z)}{G}.
	\label{eq:4.5}
\end{gather}

\item Big Bang nucleosynthesis (BBN)

We take a Gaussian prior on the baryonic matter abundance, from Big Bang Nucleosynthesis measurements, of $\omega_b\equiv\Omega_b h^2=0.02235\pm0.00016$ \cite{Cooke:2017cwo}.

\item Redshift space distortions (RSD)

Finally, we also use measurements of the product $f\sigma_8$, coming from redshift space distortions \cite{Song:2008qt}. We choose the data set listed in \cite{Sagredo:2018ahx}, whose robustness has been checked, and it is composed of 22 points, which come from WiggleZ \cite{Blake:2012pj}, SDSS \cite{Zhao:2018gvb}, while the remaining 15 are from different observations, which can be taken as uncorrelated points \cite{Davis:2010sw,Feix:2015dla,Howlett:2014opa,Song:2008qt,Blake:2013nif,Samushia:2011cs,Sanchez:2013tga,Chuang:2013wga,Huterer:2016uyq,Turnbull:2011ty,Hudson:2012gt,Pezzotta:2016gbo,Okumura:2015lvp}. We also remember to consider the model dependency of these measurements, meaning that we need to take into account the fiducial value of $\Omega_m$ used for each point. Consequently, the theoretical $f\sigma_8$ quantity is re-expressed as $f\sigma_8/\left[\frac{D_A(z,\Omega_{m})H(z,\Omega_{m})}{D_A(z,\Omega_{m,fid})H(z,\Omega_{m,fid})}\right]$, where $D_A(z)$ is the angular distance.

\end{itemize}

\begin{table*}[t]
		\centering
		\scriptsize{
			\begin{tabular*}{\textwidth}{p{1.0cm} p{1.3cm} p{1.5cm} p{1.3cm} p{2.1cm} p{1.8cm} p{1.7cm} p{1.7cm}}
				\hline
				\hline
				Model  & $H_0$ \scriptsize{[Km/s/Mpc]} & $\Omega_{m,0}$ & $b$ & $\omega_b$ & $\mathcal{M}_B$ & $\sigma_8$ & $S_8$   \\ 
				\hline
				\hline
				\\			
				
				\multicolumn{8}{c}{{CC + BAO + BBN}} \\				
				\\			
				
				$\Lambda$CDM &  $67.2\pm 1.2$  & $0.348\pm 0.021$  & $-$  & $0.02235\pm 0.00016$& $-$ & $-$ & $-$  \\
				
				$f(T)$ &  $70.3^{+1.4}_{-2.0}$  & $0.343\pm 0.020$  & $0.87^{+0.20}_{-0.43}$  & $0.02235\pm 0.00016$& $-$ & $-$ & $-$  \\	

				\\						
				\multicolumn{8}{c}{{CC + BAO + BBN + SN}} \\				
				\\			
				
				$\Lambda$CDM &  $67.0\pm 1.1 $  & $0.340\pm 0.014$  & $-$  & $0.02235\pm 0.00016$& $-19.448\pm 0.037$ & $-$ & $-$  \\
				
				$f(T)$ &  $69.3\pm 1.1$  & $0.341\pm 0.020$  & $0.746^{+0.049}_{-0.061}$  & $0.02235\pm 0.00016$& $-19.327\pm 0.079$ & $-$ & $-$  \\

				\\
				\multicolumn{8}{c}{{CC + BAO + BBN + RSD}} \\				
				\\			
				
				$\Lambda$CDM & $66.8\pm 1.2$  & $0.335\pm 0.020$  & $-$  & $0.02236\pm 0.00016$& $-$ & $0.745\pm 0.030$ & $0.786\pm 0.030$  \\
				
				$f(T)$ &  $69.8^{+1.3}_{-1.8}$  & $0.338\pm 0.020$  & $0.668^{+0.084}_{-0.30}$  & $0.02235\pm 0.00016$& $-$ & $0.779^{+0.052}_{-0.066}$ & $0.827^{+0.059}_{-0.081}$ \\
				
				\\
				\multicolumn{8}{c}{{CC + BAO + BBN + SN + RSD}} \\				
				\\			
				
				$\Lambda$CDM & $66.8\pm 1.1$  & $0.334\pm 0.013 $  & $-$  & $0.02236\pm 0.00016$& $-19.455\pm 0.037$ & $0.745\pm 0.029$ & $0.786\pm 0.030$  \\
				
				$f(T)$ &  $69.1\pm 1.1$  & $0.333\pm 0.019$  & $0.757^{+0.049}_{-0.063}$  & $0.02235\pm 0.00016$& $-19.346\pm 0.078$ & $0.745\pm 0.029$ & $0.785\pm 0.035$ \\
				
			    \\
						
				\hline
				\hline

		\end{tabular*}}
		\caption{Cosmological constraints for both models investigated at $68\%$ confidence level. We divide the results into CC+BAO+BBN, CC+BAO+BBN+SN, CC+BAO+BBN+RSD and CC+BAO+BBN+SN+RSD data sets.}
		\label{table:1}
\end{table*}

We use the emcee sampler \cite{Foreman-Mackey:2012any} to perform the statistical analysis, and the GetDist \cite{Lewis:2019xzd} Python module to analyze and plot the resulting chains. To compare the models analyzed, we use the Akaike Information Criterion (AIC) and the Bayesian Information Criterion (BIC), defined respectively as \cite{Liddle:2007fy} 

\begin{align}
AIC&\equiv -2\ln \mathcal{L}_{max} + \frac{2k(k+1)}{N-k-1},\\
BIC&\equiv -2\ln \mathcal{L}_{max} + k\ln N,\\
\label{eq:4.6}
\end{align}
where $\mathcal{L}_{max}$ is the value of the maximum likelihood, while $k$ and $N$ are the number of free parameters of the model and the total number of data points, respectively. For these criteria, the more free parameters a model has, the less preferable it becomes according to data. Defining $\Delta IC\equiv IC_{model}-IC_{ref}$, with $IC_{ref}$ representing the AIC/BIC of the reference model, that we take as being the $\Lambda$CDM one. To facilitate the comparison with previous works, we follow \cite{Anagnostopoulos:2019miu,Liddle:2007fy}, where $\Delta IC\leq 2$ corresponds to statistical compatibility, $2<\Delta IC<6$ represents a significant tension, $6 < \Delta IC < 10$, represents a strong tension, and $\Delta IC>10$ represents a very strong tension between models. As for the data set combinations considered, we establish CC+BAO+BBN as the baseline, in a way that we add the SN and RSD datasets separately, and then perform a joint analysis with all data. 

\begin{table*}[t]
	\centering
	\scriptsize{
		\begin{tabular}{c c c c c c}
			\hline
			\hline
			Model & $\chi^2_{min}$ & AIC & $\Delta$AIC & BIC & $\Delta$BIC  \\
			\hline
			\hline
			
			\\								
			\multicolumn{6}{c}{{CC + BAO + BBN}} \\				
			\\			
			
			$\Lambda$CDM & $19.462$ & $26.077$  & $0$  & $30.745$ & $0$    \\
			
			$f(T)$ & $20.062$ &  $29.115$  & $3.038$  & $35.107$ & $4.362$   \\
			
			\\				
			\multicolumn{6}{c}{{CC + BAO + BBN + SN}} \\				
			\\			
			
			$\Lambda$CDM & $1476.684$ &  $1484.708$  & $0$  & $1506.359$ & $0$   \\
			
			$f(T)$ & $1476.013$ &  $1486.049$  & $1.341$  & $1513.107$ & $6.748$   \\
			
			\\
			\multicolumn{6}{c}{{CC + BAO + BBN + RSD}} \\				
			\\			
			
			$\Lambda$CDM & $34.594$ & $43.261$  & $0$  & $51.292$ & $0$ \\
			
			$f(T)$ & $33.283$ &  $44.300$  & $1.039$  & $54.155$ & $2.863$   \\
			
			\\
			\multicolumn{6}{c}{{CC + BAO + BBN + SN + RSD}} \\				
			\\			
			
			$\Lambda$CDM & $1491.573$ & $1501.608$  & $0$  & $1528.732$ & $0$   \\
			
			$f(T)$ & $1490.491$  & $1502.541$ &  $0.933$ & $1535.082$  & $6.350$   \\

			\\		
			\hline
			\hline

	\end{tabular}}
	\caption{Statistical criteria for both $\Lambda$CDM and $f(T)$ models, where we have considered the data set combinations as in Table \ref{table:1}.}
	\label{table:2}
\end{table*}

\section{Results}\label{sec5}

\begin{figure*}
	\centering
	\includegraphics[width=0.7\columnwidth]{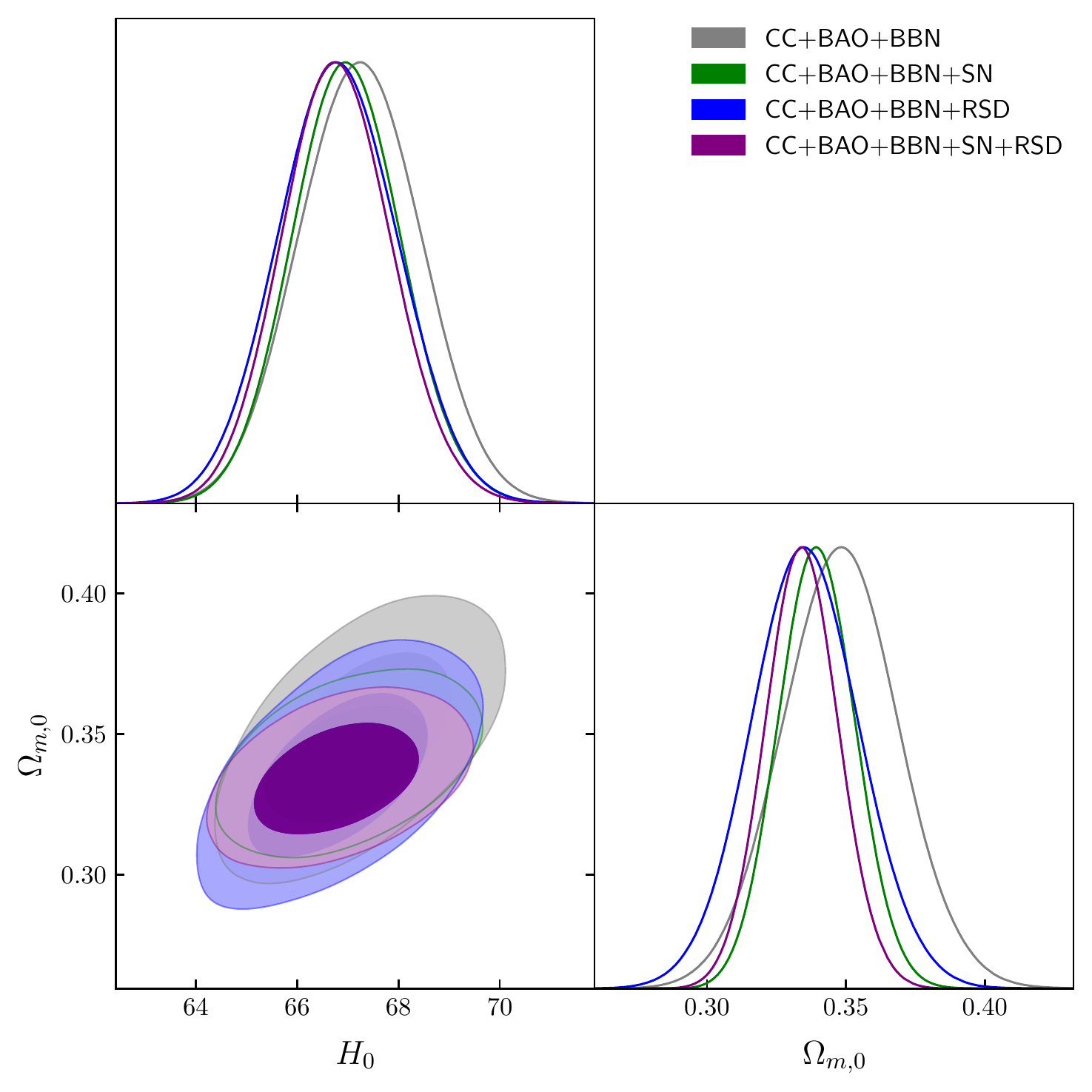}	
	\caption{Confidence contours and posteriors for the $\Lambda$CDM model, for CC+BAO+BBN (gray), CC+BAO+BBN+SN (green), CC+BAO+BBN+RSD (blue) and CC+BAO+BBN+SN+RSD (purple) data set combinations.}
	\label{fig:3}
\end{figure*}

\begin{figure*}
	\centering
	\includegraphics[width=\columnwidth]{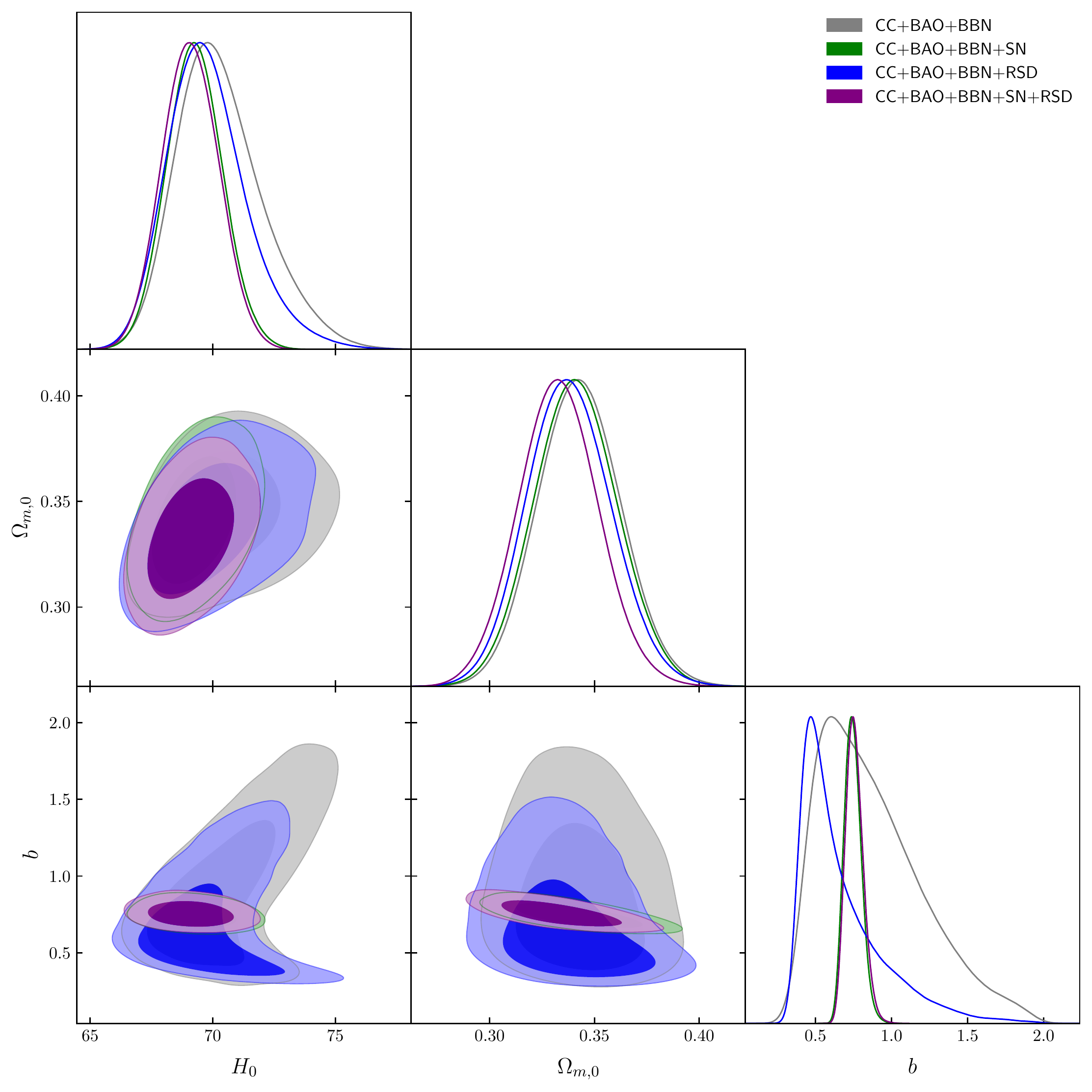}	
	\caption{Confidence contours and posteriors for the generalized exponential $f(T)$ model, where the contour and posterior colors follow fig. \ref{fig:3}.}
	\label{fig:4}
\end{figure*}

\begin{figure*}
	\includegraphics[width=0.5\columnwidth]{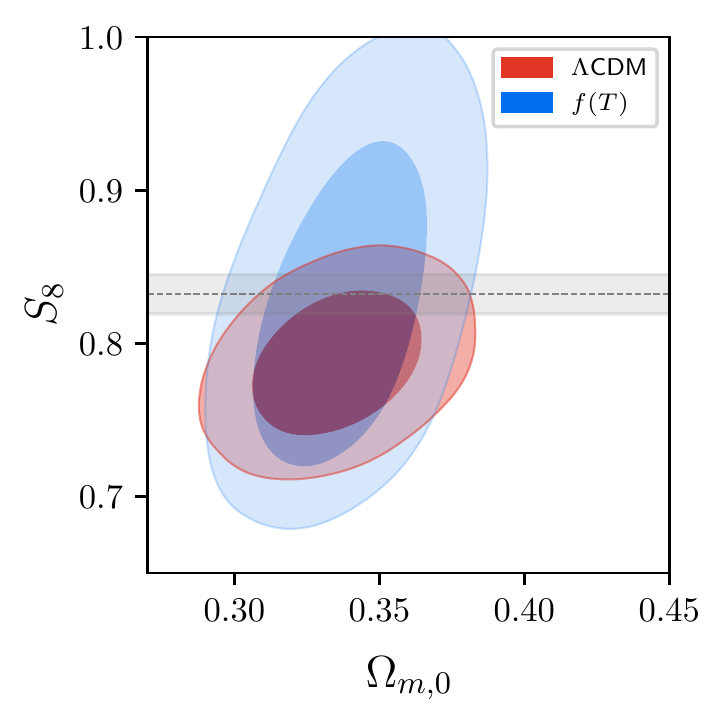}	
	\includegraphics[width=0.5\columnwidth]{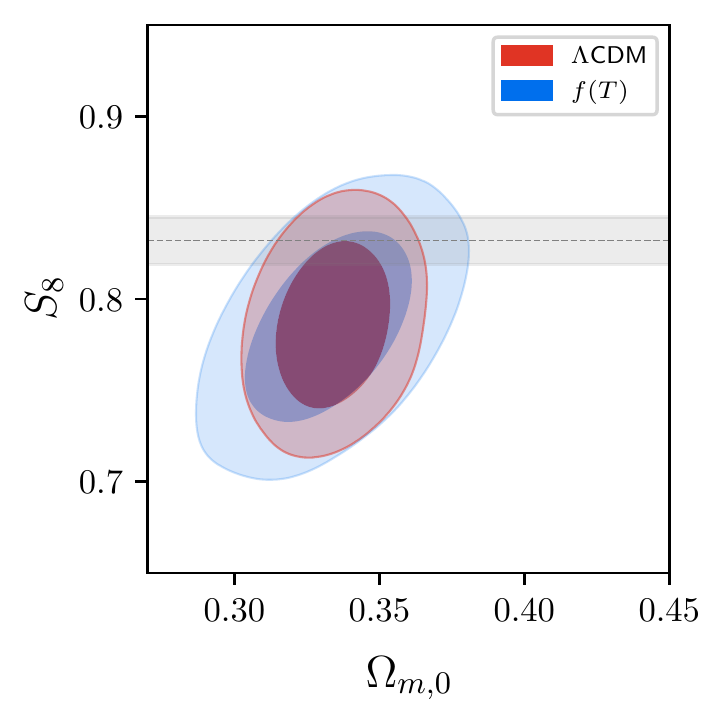}		
	\caption{The $\Omega_{m,0}-S_8$ plots for both $\Lambda$CDM and generalized exponential $f(T)$ models, for CC+BAO+BBN+RSD (left panel) and CC+BAO+BBN+SN+RSD (right panel). The black dashed line represents the Planck estimate, of $S_8=0.832\pm 0.013$, with the gray horizontal band limiting the $1\sigma$ region.}
	\label{fig:5}
\end{figure*}

\begin{figure*}
	\includegraphics[width=0.5\columnwidth]{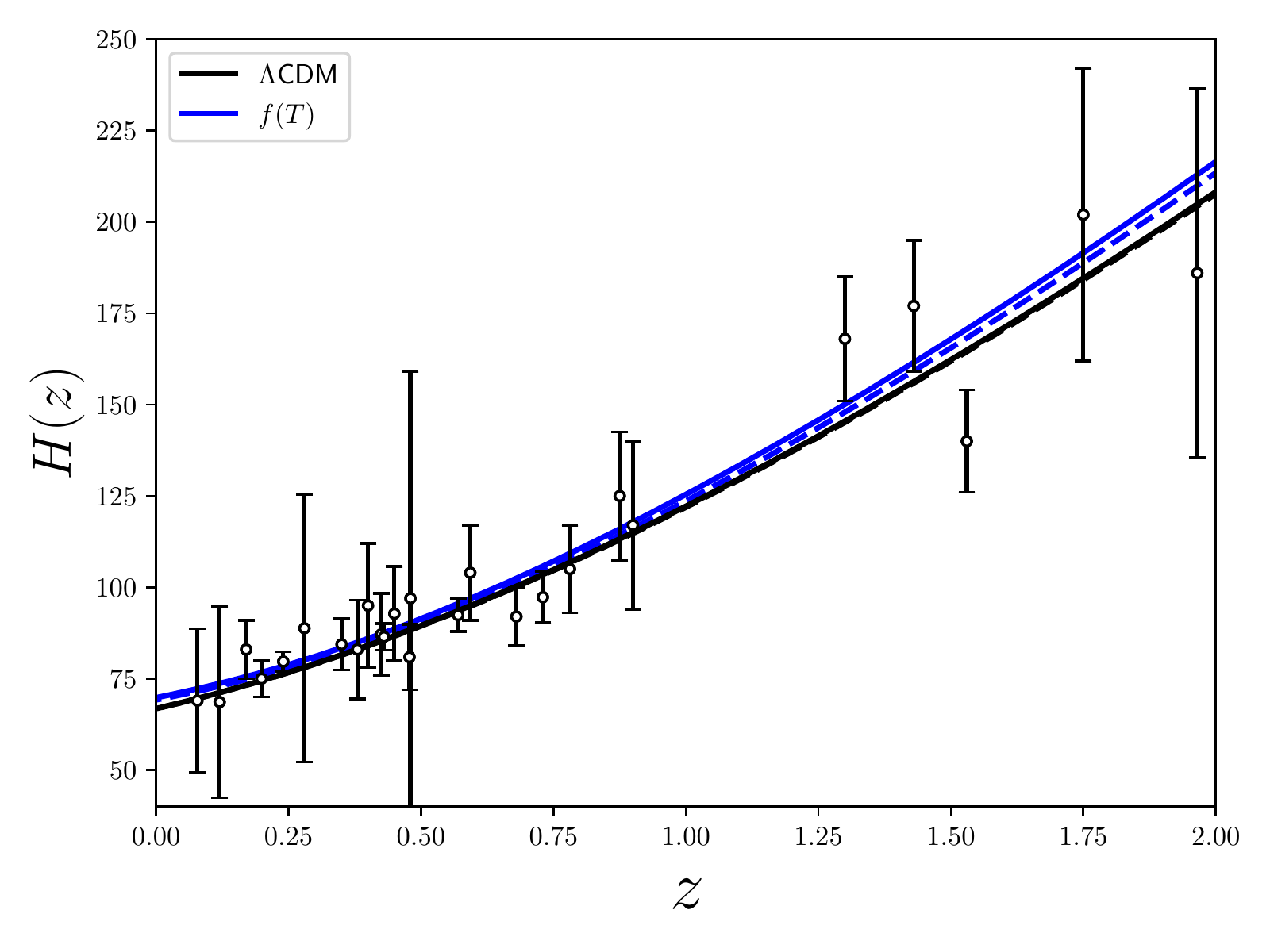}	
	\includegraphics[width=0.5\columnwidth]{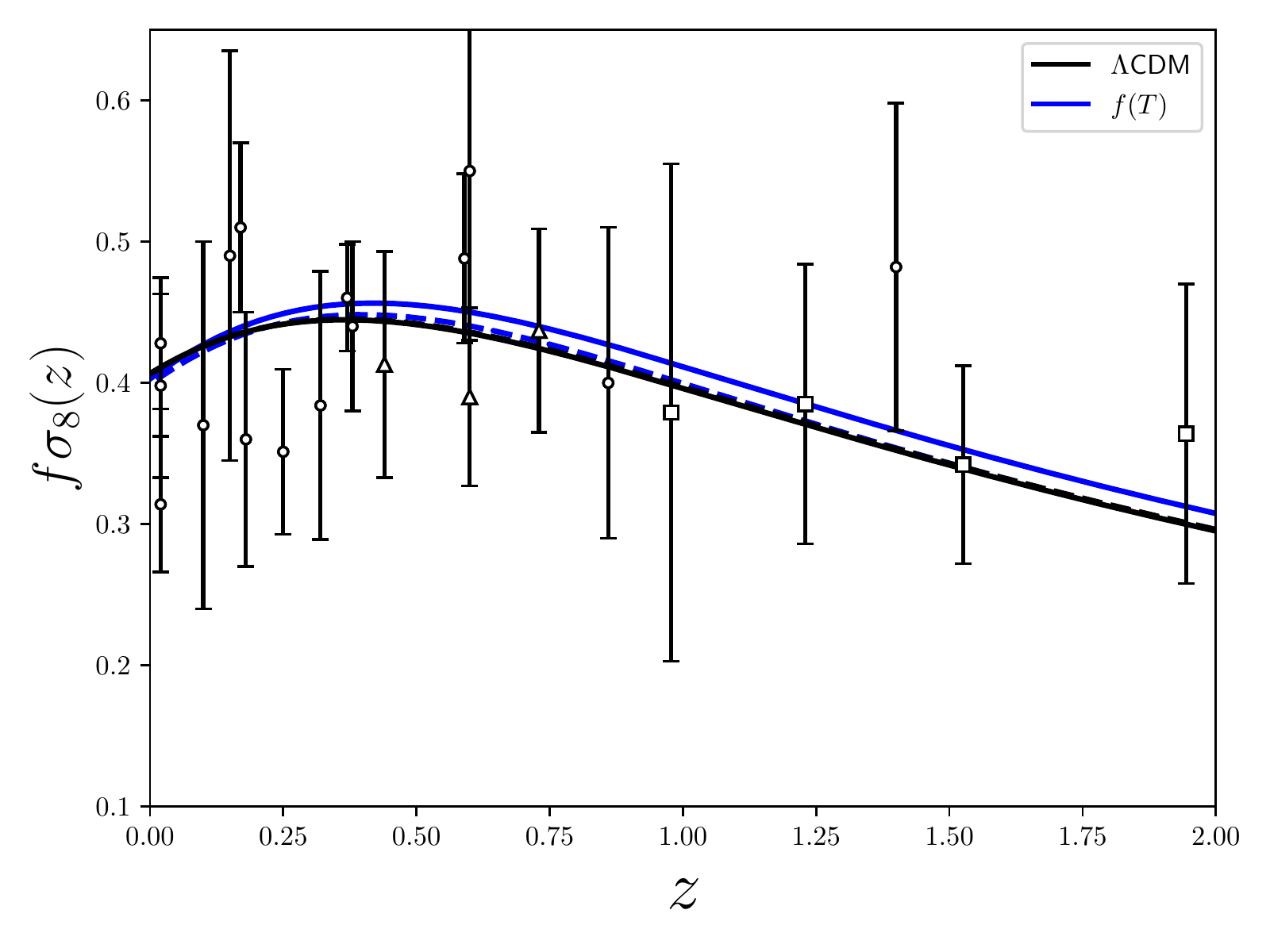}	
	\caption{Best fit curves for the Hubble parameter (left panel) and the $f\sigma_8$ quantity (right panel), for the $\Lambda$CDM model (black) and the generalized exponential $f(T)$ model (blue). Solid lines correspond to the CC+BAO+BBN+RSD results, while dashed lines correspond to the CC+BAO+BBN+SN+RSD analysis results.}
	\label{fig:6}
\end{figure*}

We then obtain restrictions on the free parameters for both $\Lambda$CDM and modified exponential $f(T)$ models for the data set combinations considered. The estimates at 68\% confidence level (CL) are shown in Tables \ref{table:1}, \ref{table:2}, while the respective confidence contours are shown in figures \ref{fig:3}, \ref{fig:4}, \ref{fig:5} and \ref{fig:6}. For the standard $\Lambda$CDM model, we find constraints that are compatible with the existing literature. The baseline combination CC+BAO+BBN leads to a value for $H_0$ of $H_0=67.2\pm 1.2$ Km/s/Mpc, while the total matter fraction is estimated as $\Omega_{m,0}=0.348\pm0.021$, a result higher than the Planck estimate, but still compatible at $2\sigma$ level. Next, we compare these results with the ones obtained for the $f(T)$ model. We note that the value of the present Hubble parameter is higher, of $H_0=70.3^{+1.4}_{-2.0}$ Km/s/Mpc, with a higher uncertainty when compared to the standard model estimate. Consequently, we see that the $H_0$ tension seems to be alleviated within the $f(T)$ model, as we obtain $\sim 1.5\sigma$ when our estimate is compared with the late-time one from SH0ES. A similar fraction of matter to that of the $\Lambda$CDM model is also obtained, being $\Omega_{m,0}=0.343\pm 0.020$. The $f(T)$ function parameter $b$, which characterizes the deviation from general relativity, is constrained at $b=0.87^{+0.20}_{-0.43}$, already indicating a phantom crossing of $\omega_{DE}$ at present times. Fig. \ref{fig:4} however, shows that values where $b<0.5$ are not excluded as well as the $b=1$ case. Adding SN data from Pantheon+ significantly improves the constraints. For the CC+BAO+BBN+SN data set combination, we obtain $H_0=69.3\pm 1.1$ Km/s/Mpc for the generalized exponential $f(T)$ model, a value that still alleviates the tension when compared with the $\Lambda$CDM estimate, especially, when we note that the $1\sigma$ uncertainty for the parameter is now the same for both models. Moreover, it is now possible to obtain much more restrictive constraints on the deviation parameter $b$, being estimated as $b=0.746^{+0.049}_{-0.061}$, excluding the $b=1$ case at more than $2\sigma$ confidence level, while also implying a phantom crossing of the dark energy equation of state at around $z_{tr}\sim 1$. Also, the problematic region where the positivity of $\rho_{DE}$ is violated is excluded as well.

Next, we include information on the growth of large structures, characterized by the RSD measurements. We obtain a value of $\sigma_8=0.779^{+0.052}_{-0.066}$ for the $f(T)$ model, compatible with the $\Lambda$CDM estimates previously achieved in the literature \cite{Sagredo:2018rvc,Sagredo:2018ahx,Quelle:2019vam}, but higher than the one obtained here, of $\sigma_8=0.745\pm 0.030$. As a result, we obtain a value that does not worsen the tension, although we should note that the uncertainty is significantly larger. This compatibility with the standard model is more evident by the inclusion of SN data for the joint analysis, where both models yield almost identical estimates. Furthermore, the tension between early and late-time estimates can be expressed by the $S_8$ quantity. We remember that Planck infers a value of $S_8=0.832\pm 0.013$, for the $\Lambda$CDM model; on the other hand, a variety of low-redshift measurements seem to imply the existence of a tension for the growth of structures \cite{Abdalla:2022yfr}, such as the Kilo-Degree Survey (KiDS) + VIKING determination from cosmic shear, of $S_8=0.737^{+0.040}_{-0.036}$ \cite{Hildebrandt:2018yau}, and the latest Dark Energy Survey (DES-Y3), of $S_8=0.776\pm0.017$ \cite{DES:2021wwk}. We have checked the initial impact of the extended scenario in the determination of $S_8$. For the CC+BAO+BBN+RSD data set, while we have obtained $S_8=0.786\pm 0.030$ for the standard $\Lambda$CDM model, the generalized exponential $f(T)$ one seems to prefer a higher value of $S_8=0.827^{+0.059}_{-0.081}$, initially solving the tension with the Planck estimate. A better view of the results is shown in fig. \ref{fig:5} (left panel), where the $\Omega_{m,0}-S_8$ plane for both models indicate that the larger uncertainty in the estimate for the exponential $f(T)$ allows for compatibility with the Planck determination, although the $\Lambda$CDM estimate does not present a strong incompatibility with Planck, according to the considered data. A similar conclusion was achieved in \cite{Nunes:2021ipq}, in which a joint analysis leads to a lower level of disagreement between estimates within the $\Lambda$CDM model. On the other hand, as mentioned, the inclusion of SN data provides a tighter constraint on the other models, which impacts the confidence contours on the $\Omega_{m,0}-S_8$ estimates as well. Also in Figure \ref{fig:5}, it is possible to see the equivalence of both models when all data is considered, such that there is no worsening of the growth tension within the $f(T)$ model.

Focusing now on the deviation parameter $b$, it is easy to see that the inclusion of SN data (see fig. \ref{fig:4}) allows for a precise estimate, reflecting directly on the equation of state parameter behavior, as well as the evolution of the effective gravitational constant, as discussed. Both CC+BAO+BBN+SN and CC+BAO+BBN+SN+RSD give similar constraints, with the latter data set giving $b=0.757^{+0.049}_{-0.063}$. We note that all estimates agree to a value lower than one; this has the consequence of a compatibility with a phantom crossing within the model, and leading to a lower effective gravitational constant at present (fig. \ref{fig:1}, lower panel and fig. \ref{fig:2}, left panel). Another point is that the inclusion of SN data excludes the parameter space where $b<0.5$ by more than $2\sigma$ confidence level, therefore not including the problematic feature where a negative $\Omega_{DE}$ is present at some time in cosmic evolution. In Figure \ref{fig:6}, the best fit curves for the Hubble parameter (left panel) and $f\sigma_8$ (right panel) are shown, where one notes the difference between the two models, where in particular, the impact of the $f(T)$ scenario is more noticeable in the $H(z)$ behavior, whereas the $f\sigma_8$ quantity closely matches that of the $\Lambda$CDM one, for the joint analysis results.

Finally, it is also useful to look at the preference of the data for the models considered. This can be done by computing the statistical criteria, such as AIC/BIC, discussed in the previous section. In Table \ref{table:2}, we show the presented criteria for both models and all data set combinations considered. For the analysis involving CC+BAO+BBN data, while we see that the AIC/BIC indicate a slight tension between the models, the situation changes when SN data is included. We note that the tension according to AIC decreases, while BIC indicates a stronger preference for the standard model. We also note that as we consider other data set combinations, the $f(T)$ becomes more competitive with respect to the $\Lambda$CDM one. For CC+BAO+BBN+RSD, AIC/BIC seem to prefer the $\Lambda$CDM model, but a statistical equivalence can be achieved. Finally, for the joint analysis with all data sets, we are able to achieve the lowest tension between models, according to the AIC. These results indicate that even with the presence of an extra parameter, the $f(T)$ scenario is competitive when compared with the standard $\Lambda$CDM model, a feature already discussed in \cite{Xu:2018npu,Anagnostopoulos:2019miu}, for other $f(T)$ models.

\section{Conclusions}\label{sec6}

$f(T)$ gravity has been widely considered in the last years as a promising alternative for a description of gravity and late-time phenomena. Through a function that expresses a deviation from the standard $\Lambda$CDM model, it is possible to estimate how much of a modification is necessary and preferred by data, in order to have a correct cosmic history, at the same time that one can address current problems, such as the Hubble tension. Furthermore, these classes of models have important implications at the cosmological perturbations level as well \cite{Benetti:2020hxp,Hashim:2021pkq}, at the same time that the impact from structure formation at the non-linear level is also starting to be investigated \cite{Malekjani:2016mtm,Lin:2016xyi,Huang:2022slc}. A variety of functional forms for the $f(T)$ function have been discussed and compared with data, in a manner that these results can lead us to well motivated forms that can lead to a more unified picture that offers a good description of the universe at both early and late times. 

In this work, we have investigated a $f(T)$ model which allows a phantom crossing of the equation of state parameter $\omega_{DE}$. The model can be seen as a generalization of the scenario investigated in \cite{Awad:2017yod,Hashim:2020sez,Hashim:2021pkq}, in a manner that it is possible to check for a deviation of this model, resulting in an interesting phenomenology, as well as discuss the issue of cosmic tensions. The model represents a modification from GR, in the sense that it cannot reproduce the exact $\Lambda$CDM cosmology for a choice of parameters, so one can determine the level of deviation from general relativity even for the $b\neq 1$ cases. After discussing the main features of the model, we have constrained its free parameters, by using recent and robust observational data, such as the newest Type Ia supernovae measurements from Pantheon+, which adds to the previous Pantheon set, including several measurements at low redshifts. Additionally, we have considered the effects of an effective gravitational constant in the distance modulus, as recently done for the first time in $f(T)$ gravity in \cite{Kumar:2022nvf}. As a result, we have found a significant restriction on the deviation parameter $b$, indicating a phantom crossing, according to the model, for all data sets considered, while being competitive with the $\Lambda$CDM model through statistical criteria. Moreover, we also see that especially the Hubble tension seems to be alleviated, as well as a non-worsening of the growth tension, which motivates a deeper investigation of these types of scenarios. The next step then, is to consider for instance the full Planck likelihood, which will give clearer answers on the $S_8$ estimate, as well as an understanding on the consistency of this scenario for early-time physics. These questions will be discussed in a future work.

\section*{Acknowledgments}

F.B.M. dos Santos is supported by Coordenação de Aperfeiçoamento de Pessoal de
Nível Superior (CAPES). This work was developed thanks to the High Performance Computing Center at the Universidade Federal do Rio Grande do Norte (NPAD/UFRN). The Author is thankful to Micol Benetti for useful discussions. 

\bibliography{references}

\end{document}